\documentclass[12pt]{article}
\usepackage{enumitem}
\usepackage{dsfont}
\usepackage{mathrsfs}
\usepackage{natbib,graphicx,setspace,lscape,longtable}
\usepackage{mathrsfs,amsmath,amsthm,amssymb,color}
\usepackage{natbib,epsfig,graphicx,pdfpages}
\usepackage{rotating}
\usepackage{float}
\usepackage{bm}
\usepackage{ulem}
\usepackage{CJKutf8}
\usepackage{algorithm}
\usepackage{algorithmic}
\usepackage{booktabs}
\usepackage{bm}
\usepackage{subfigure}
\usepackage{makecell}
\usepackage{hyperref}
\usepackage{multirow}
\usepackage{threeparttable}
\usepackage{xr}
\usepackage[dvipsnames]{xcolor}
\usepackage{mathtools}
\usepackage[title]{appendix}
\hypersetup{
	colorlinks=true,
	linkcolor=blue,
	citecolor=blue,
	urlcolor=blue}

\bibpunct{(}{)}{;}{a}{,}{,}

\newtheorem{theorem}{Theorem}
\newtheorem{lemma}{Lemma}
\newtheorem{proposition}{Proposition}

\newcommand{\csection}[1]
{\refstepcounter{section}
	\begin{center}
		{\bf\large\arabic{section}. #1}
	\end{center}
}

\newcommand{\csubsection}[1]{
	\refstepcounter{subsection}
	\begin{center}
		{\it\arabic{section}.\arabic{subsection}. #1}
		
	\end{center}
}

\def\tr{\mbox{tr}}

\def\tfmle{\textup{fmle}}

\def\beq{\begin{equation}}
	\def\eeq{\end{equation}}
\def\beqr{\begin{eqnarray}}
	\def\eeqr{\end{eqnarray}}
\def\beqrs{\begin{eqnarray*}}
	\def\eeqrs{\end{eqnarray*}}
\def\bet{\begin{theorem}}
	\def\eet{\end{theorem}}
\def\bel{\begin{lemma}}
	\def\eel{\end{lemma}}
\def\bes{\begin{step}}
	\def\ees{\end{step}}
\def\bep{\begin{proposition}}
	\def\eep{\end{proposition}}
\def\bg{\begin{figure}[tbph]\begin{center}}
		\def\eg{\end{center}\end{figure}}

\def\bc{\begin{center}}
	\def\ec{\end{center}}

\def\tcmle{\textup{cmle}}

\def\wt{\widetilde}

\def\wh{\widehat}

\def\mA{\mathcal A}

\def\mD{\mathcal D}
\def\mQ{\mathcal Q}
\def\mR{\mathbb{R}}
\def\mS{\mathcal S}

\def\mL{\mathcal L}
\def\mB{\mathcal B}

\def\1{\mbox{\boldmath $1$}}

\def\bE{\mathbb E}

\def\mX{\mathbb{X}}

\def\mY{\mathbb{Y}}
\def\mZ{\mathbb{Z}}
\def\var{\mbox{var}}

\def\cov{\mbox{cov}}
\def\tr{\mbox{tr}}
\def\argmin{\mbox{argmin}}
\def\argmax{\mbox{argmax}}
\def\rank{\mbox{rank}}
\def\diag{\mbox{diag}}

\definecolor{darkgreen}{RGB}{0, 100, 0}

\newcommand{\mE}{{\mathcal E}}
\numberwithin{equation}{section}
\makeatletter

\newcommand{\Rmnum}[1]{\expandafter\@slowromancap\romannumeral #1@}
\makeatother

\begin{document}
\begin{CJK}{UTF8}{gbsn}
\begin{center}
  {\bf\Large High-Dimensional Spatial Autoregression with Latent Factors by Diversified Projections}\\
		\bigskip
Jiaxin Shi$^1$, Xuening Zhu$^{2,*}$, Jing Zhou$^3$, Baichen Yu$^1$, and Hansheng Wang$^1$

{\it $^1$Guanghua School of Management, Peking University, Beijing, China}\\
{\it $^2$School of Data Science, Fudan University, Shanghai, China}\\
{\it $^3$Center for Applied Statistics, School of Statistics, Renmin University of China, Beijing, China}
\end{center}
\begin{footnotetext}[1]
{Xuening Zhu is the corresponding author.
}
\end{footnotetext}
\begin{singlespace}
\begin{abstract}

We study one particular type of multivariate spatial autoregression (MSAR) model with diverging dimensions in both responses and covariates. This makes the usual MSAR models no longer applicable due to the high computational cost. To address this issue, we propose a factor-augmented spatial autoregression (FSAR) model. FSAR is a special case of MSAR but with a novel factor structure imposed on the high-dimensional  random error vector. The latent factors of FSAR are assumed to be of a fixed dimension. Therefore, they can be estimated consistently by the diversified projections method \citep{fan2022learning}, as long as the dimension of the multivariate response is diverging. Once the fixed-dimensional latent factors are consistently estimated, they are then fed back into the original SAR model and serve as exogenous covariates. This leads to a novel FSAR model. Thereafter, different components of the high-dimensional response can be modeled separately. To handle the high-dimensional feature, a smoothly clipped absolute deviation (SCAD) type penalized estimator is developed for each response component. We show theoretically that the resulting SCAD estimator is uniformly selection consistent, as long as the tuning parameter is selected appropriately. For practical selection of the tuning parameter, a novel BIC method is developed. Extensive numerical studies are conducted to demonstrate the finite sample performance of the proposed method.

\vspace{1em}
				
\end{abstract}

\end{singlespace}

\newpage
\csection{INTRODUCTION}\label{sec_intro}

Spatial data are frequently encountered in various statistical and econometric applications \citep{fujita2001spatial,yin2022spatial,zhou2023spatial}. These applications include, but are not limited to, environmental analysis \citep{zhou2023spatial}, geographical science \citep{yin2022spatial}, political economics \citep{2016strategic}, and many others. To model the spatial dependence among multiple subjects/nodes, a variety of spatial autoregressive (SAR) models have been developed and extensively studied \citep{1998generalized,lee2010estimation,huang2019least}. A number of estimation methods have been proposed, and their corresponding statistical properties have been carefully studied. Those estimation methods include, the quasi-maximum likelihood estimation (QMLE) method \citep{lee2004asymptotic}, the generalized method of moments (GMM) \citep{lee2007gmm}, the least squares estimation (LSE) methods \citep{huang2019least}, and many others \citep{su2012semiparametric}.

It is remarkable that those classical SAR models are applicable only to datasets with univariate responses. In other words, only a univariate response is collected for every subject/node in a spatial/network dataset. However, in real-world applications, multivariate responses are frequently encountered. This leads to various multivariate spatial autoregressive (MSAR) models \citep{yang2017identification,zhu2020multivariate}. Moreover, real network datasets with high-dimensional multivariate responses are becoming increasingly available \citep{zhang2022joint,chen2025estimating}. Consider, for example, a regional economic dataset from China. The full dataset contains a total of 287 cities and 112 macroeconomic indicators. The objective here is to study the spatial spillover effects of regional economics, which is a problem of great importance for understanding spatial economic dynamics and regional economic growth in macroeconomic research \citep{anselin1988spatial,zhou2023spatial}. For this dataset, each region can be treated as a node, which is spatially connected with others. This makes the SAR model a natural choice in empirical economics literature  \citep{blasques2016spillover,de2024identifying}. In addition to the spatial structure, we also observe a large number of economic indicators for each region (i.e., node). This leads to a high-dimensional response vector for each region. As a consequence, the MSAR models of \cite{yang2017identification} and \cite{zhu2020multivariate} are difficult to apply directly. This is mainly because the associated computational cost becomes extremely expensive if the response dimension is relatively high. This interesting dataset motivates us to develop a novel method, which is able to model spatial dependence for datasets with high-dimensional responses.

To this end, we propose a factor-augmented SAR approach. This method assumes a standard SAR model for each component of the high-dimensional response. By doing so, the spatial dependence structure can be flexibly modeled for different response components in a parallel way. This leads to a high-dimensional error vector for each node. To analyze the high-dimensional data, various factor modeling techniques have been developed \citep{fan2008high,bai2012statistical,lam2012factor,fan2022learning}. We are then inspired to impose a factor model on this high-dimensional error vector. This leads to a new type of SAR model with a factor-augmented structure. For convenience, we refer to this as a factor-augmented spatial autoregressive (FSAR) model. It is worth noting that the FSAR model is related to the dynamic SAR model in the existing literature \citep{bai2021dynamic}. However, there are two critical differences. First, FSAR is a static MSAR model without a dynamic panel structure over time. Second, FSAR allows different spatial effects for different responses. By assuming that the factor dimension is fixed as $d$, we obtain a highly simplified model structure with a total of only $\big(d+q+2\big)p$ parameters, if the exogenous covariates are of dimension $q$. In contrast, a traditional MSAR model in this case \citep{yang2017identification,zhu2020multivariate} should consume a total of $(3p^2/2+pq)$ parameters. Moreover, it is remarkable that the type of dependence captured by MSAR and FSAR models are different. The MSAR model is good at capturing weak dependence, which refers to the type of influence due to local network neighbors. In contrast, our FSAR model is good at capturing strong dependence, which reflects the type of the influence due to the global cross-sectional dependence. Therefore, the applications of MSAR and FSAR models are not the same.

To practically estimate the FSAR model, a three-step estimation procedure is developed. In the first step, the standard QMLE method \citep{lee2004asymptotic} is applied to each component of the high-dimensional response. This yields to a consistent initial estimator for each componentwise SAR model. Accordingly, the high-dimensional error vector can be consistently differentiated for each node. In the second step, the diversified projections method of \cite{fan2022learning} is applied to the estimated error vectors for factor estimation. By doing so, the latent factors can be consistently estimated up to an affine transformation. In the last step, we treat the estimated factors as exogenous covariates. Then, the FSAR model can be estimated for each response component in a fully parallel way \citep{lee2004asymptotic,lee2010estimation}. This leads to the final estimators for the spatial correlation parameters. Under appropriate regularity conditions, we show theoretically that the resulting estimator is $\sqrt{n}$-consistent and asymptotically normal. To handle high-dimensional exogenous covariates, a smoothly clipped absolute deviation (SCAD) penalized estimator \citep{fan2001variable} is developed for the FSAR model, and a novel BIC method is developed for tuning parameter selection \citep{wang2007tuning,chen2008extended,wang2009shrinkageB}. We show theoretically that the resulting estimator is uniformly selection consistent for every response component. 

The rest of the article is organized as follows. Section~\ref{sec_meth} develops the FSAR model. The estimation methods and the associated asymptotic theory are also included. The numerical studies are presented in Section~\ref{sec_numer}, which includes both extensive simulation experiments and a real data example. Finally, Section~\ref{sec_conclu} concludes the article with a brief discussion. All technical proofs are left to the Appendix.

\csection{METHODOLOGY}\label{sec_meth}
\csubsection{The Model Setup}\label{subsec_setup}

Consider a large-scale network with $n$ nodes indexed by $1\leq i\leq n$. Define an adjacency matrix of the network as $A = (a_{i_1i_2})\in\mR^{n\times n}$, where $a_{i_1i_2} = 1$ if the node $i_1$ is connected to the node $i_2$ and $a_{i_1i_2} = 0$ otherwise. Following the existing literature \citep{lee2004asymptotic,zhu2020multivariate}, we set $a_{ii}=0$ for every $1\leq i\leq n$. Next, define a spatial weight matrix as $W = (w_{i_1i_2})\in\mR^{n\times n}$ with $w_{i_1i_2} = a_{i_1i_2}/ n_{i_1}$ and $n_{i_1} = \sum_{i_2=1}^na_{i_1i_2}$ so that each row of the weight matrix $W$ sums up to one. For those zero-degree nodes (i.e., $\sum_{i_2=1}^n a_{i_1i_2} = 0$), we set $w_{i_1i_2}=0$ for $1\leq i_2\leq n$ so that the useful covariate information contained in those nodes can be maintained. Next, for each node $i$, we observe a $p$-dimensional response vector $Y_i=(Y_{ij})\in\mR^p$ with $p\to\infty$ as $n\to\infty$ and a $q$-dimensional exogenous covariate vector as $X_i = (X_{im})\in\mR^q$ \citep{lee2004asymptotic,lee2010estimation,huang2021feature}. Write $\mY_j = (Y_{ij})\in\mR^n$ as the response vector for the $j$-th component. Accordingly, write $\mX = (X_{1},\ldots,X_{n})^{\top}\in\mR^{n\times q}$ as the covariate matrix. Each component of $\mY_j$ is expected to be spatially correlated with others through $A$. Therefore, we assume for $\mY_j$ a standard spatial autoregressive (SAR) model as
\begin{equation}\label{model-SAR}
	\mY_j = \rho_j W \mY_j + \mX\beta_j + \mE_j,
\end{equation}
where $\rho_j\in\mR$ is the spatial correlation, $\beta_j = (\beta_{jm})\in\mR^q$ is the coefficient vector, and $\mE_j = (\varepsilon_{ij})\in\mR^n$ is the error vector.

Let $\rho_j^*$ and $\beta_j^* = (\beta_{jm}^*)$ be the true values of $\rho_j$ and $\beta_j$, respectively. In this work, we allow $q$ to diverge in the sense that $q\to\infty$ as $n\to\infty$. In this case, we should expect that a large number of covariates are redundant for any given response $\mY_j$ \citep{tibshirani1996regression,fan2001variable,huang2021feature}. To reflect this phenomenon, define for every response $\mY_j$ a true model set $\mS_{(j),T}=\{1\leq k\leq q: \beta_{jk}^*\neq 0\}$ with size $ s_j = | \mS_{(j),T}|$. Define $\mS_{\text{True}}=\bigcup_{1\leq j\leq p}\mS_{(j),T}$. In this work, we assume the maximum size of the true model for every response component is upper bounded by a fixed number $m>0$ (i.e., $\max_{1\leq j\leq p}s_j \leq m$). However, we allow the total number of relevant covariates (i.e., $|\mS_{\text{True}}|$) to diverge as $n\to\infty$. This allows a diverging amount of information to be used. Next, let $\mX_k = (X_{ik})\in\mR^n$ be the $k$-th column of $\mX$. Write $\mX_{(j)} = \big(\mX_k: k\in\mS_{(j),T}\big)\in\mR^{n\times s_j} $ as the submatrix of $\mX$ corresponding to $\mS_{(j),T}$. Similarly, define $\beta_{(j)} = \big(\beta_{jk}: k\in \mS_{(j),T}\big) \in\mR^{s_j}$. Then, model (\ref{model-SAR}) becomes 
\begin{equation}\label{model-SAR-new}
	\mY_j = \rho_j W \mY_j + \mX_{(j)}\beta_{(j)} + \mE_j.
\end{equation}
Assume that $ \mS_{(j),T}s$ are already given at this moment. In practice, $ \mS_{(j),T}s$ are typically unknown. Therefore, they have to be consistently estimated based on the observed data. This is an important issue to be studied in Section~\ref{subsec_Shrink}.

Write $\varepsilon_i = (\varepsilon_{ij})\in\mR^p$ as the error vector associated with the $i$-th node. Then, how to model the stochastic behavior of $\varepsilon_i$ with a high dimension $p$ becomes a problem of great interest. To address this issue, we follow the ideas of \cite{fan2008high} and \cite{wang2012factor} and assume a factor model as 
\begin{equation}\label{equ:factor}
	\varepsilon_i = BZ_i + \omega_i,
\end{equation}
where $Z_i=(Z_{ik})\in\mR^d$ is a $d$-dimensional latent factor for the $i$-th node, $B=(b_{jk})\in\mR^{p\times d}$ is the loading matrix, and $\omega_i = (\omega_{ij})\in\mR^p$ represents the information contained in $\varepsilon_i$ but missed by $Z_i$. We assume that the factor dimension $d$ is a fixed number, consistent with the existing literature \citep{bai2012statistical,lam2012factor,fan2022learning}, and also with our empirical example, which is to be analyzed in Section~\ref{subsec_real}. We assume that $Z_i$ and $\omega_{ij}$s are mutually independent with mean $0$. Write $\cov(\varepsilon_i)=\Sigma_{\varepsilon}=(\sigma_{j_1j_2})\in\mR^{p\times p}$, $\Sigma_{\omega} = \cov(\omega_i)=(\tau_{j_1j_2})\in\mR^{p\times p}$, and $\Sigma_Z =\cov(Z_i) \in\mR^{d\times d}$. Accordingly, the true parameters are denoted as $B^* = (b_{jk}^*)$, $\Sigma_{\varepsilon}^* = (\sigma_{j_1j_2}^*)$, $\Sigma_{\omega}^* = (\tau_{j_1j_2}^*) $, and $\Sigma_Z^*$. It then follows that $\Sigma_{\varepsilon}^* = B^*\Sigma_{Z}^*B^{*\top} + \Sigma_{\omega}^*$. For model identification, we assume that $\Sigma_Z^* = I_d$, which stands for a $d$-dimensional identity matrix. Otherwise, we can always re-define $ Z_i\coloneqq \Sigma_Z^{*-1/2}Z_i$ and $ B\coloneqq B\Sigma_Z^{*1/2}$ so that model (\ref{equ:factor}) remains valid but with $\cov(Z_i) = I_d$. Let $\lambda_{\max}(A)$ and $\lambda_{\min}(A)$ be the largest and smallest eigenvalues of an arbitrary symmetric matrix $A$, respectively. Moreover, we assume that $\Sigma_{\omega}^*$ is a positive definite matrix. Notably, we do not require $\Sigma_{\omega}^*$ to be of a diagonal structure. The only constraint imposed on $\Sigma_{\omega}^*$ is that its eigenvalues are well bounded away from 0 and infinity as $p\to\infty$ \citep{wang2009shrinkageB,wang2012factor,fan2022learning}. 

\csubsection{Componentwise Maximum Likelihood Estimators}\label{subsec_CMLE}

We next consider how to estimate the model parameters. Here we temporarily assume that the true model sets $\mathcal{S}_{j,T}s$ are given. Thus, the following estimators obtained in Sections \ref{subsec_CMLE}--\ref{subsec_FMLE} are the very ideal estimators, which are often referred as the oracle estimators in the literature \citep{donoho1994ideal,fan2001variable}. Unfortunately, those true model sets $\mathcal{S}_{j,T}s$ are typically unknown in practice. Therefore, these oracle estimators cannot be practically computed and the true model sets have to be empirically estimated. That is the reason why we have developed in Section \ref{subsec_Shrink} a SCAD-penalized estimation method for a uniformly consistent selection of the true model sets $\mathcal{S}_{j,T}s$. This should be done in prior to applying the three-step estimation procedure as shown in Sections \ref{subsec_CMLE}--\ref{subsec_FMLE}.

In this study, we focus on the QMLE method due to its theoretical importance. However, the method to be developed can readily be applied to other estimation methods without additional difficulty. We then apply the QMLE method to each response component $j$ to obtain consistent initial estimators for the interested parameters (i.e., $\rho$, $\beta_{(j)}$, and $\sigma_{jj}$). For convenience, we refer to these as componentwise maximum likelihood estimators (CMLE). Specifically, under the following technical conditions \ref{con1}--\ref{con2}, we have $E(\mE_j)=0$ and $\cov(\mE_j)=\sigma_{jj}^* I_n$, where $\sigma_{jj}^* =\var(\varepsilon_{ij})= \|b_j^*\|^2+\tau_{jj}^*$ and $b_j^*$ is the $j$-th column of $B^*$. Write $ S_j=I_n-\rho_j W$. This leads to a reduced form as $\mY_j = S_j^{*-1}\big(\mX_{(j)}\beta_{(j)}^* +\mE_j\big) $ with $S_j^* = I_n - \rho_j^*W$. To ensure that $S_j$ is invertible, we follow \cite{lee2004asymptotic} and assume  $\vert\rho_j\vert<1$ for every $1\leq j\leq p$. Define $\theta_j=\big(\rho_j,\beta_{(j)}^{\top},\sigma_{jj}\big)^{\top}\in\mR^{s_j+2}$. Then, the log-likelihood function for CMLE is given by 
\begin{equation}\label{log-lik-SAR}
	\mL_{\tcmle}^{(j)}\big(\theta_j\big) = -\frac{n}{2} \log \sigma_{jj} + \log\big\vert S_j\big\vert- \frac{1}{2\sigma_{jj}}\Big(S_j \mY_j - \mX_{(j)}\beta_{(j)} \Big)^{\top}\Big(S_j \mY_j - \mX_{(j)}\beta_{(j)} \Big),
\end{equation}
where some irrelevant constants are ignored. The CMLE for $\theta_j$ can then be obtained by maximizing (\ref{log-lik-SAR}) as $\wh{\theta}_{j,\tcmle} = \big(\wh{\rho}_{j,\tcmle}, \wh{\beta}_{(j),\tcmle}^{\top}, \wh{\sigma}_{jj,\tcmle}\big)^{\top}= \argmax_{\theta}\mL^{(j)}_{\tcmle}(\theta)$. The asymptotic properties of $\widehat{\theta}_{j,\tcmle}$ have been well studied in the existing literature \citep{lee2004asymptotic,lee2010estimation,yang2017identification}. 

By Theorem 3.1 of \cite{lee2004asymptotic}, the componentwise estimator $\wh{\theta}_{j,\tcmle}$ is $\sqrt{n}$-consistent for every $1\leq j\leq p$. However, to the best of our knowledge, it seems that no uniform convergence result has been established when $p\to\infty$. Nevertheless, a uniform convergence result about $\wh{\theta}_{j,\tcmle}s$ is critically important for our subsequent theory development. To this end, define the $\ell_q$-norm of an arbitrary vector $v=(v_j)\in\mR^p$ as $\Vert v\Vert_q = \big(\sum_{j=1}^p\vert v_j\vert^q\big)^{1/q}$ for any $q>0$. For convenience, we omit the subscript $q$ when $q=2$. Additionally, denote a sub-Weibull distribution of order $\alpha$ as sub-Weibull($\alpha$). Let $U\in\mR$ be an arbitrary random variable. Define its sub-Weibull($\alpha$) norm as $ \|U\|_{\psi_\alpha} = \inf\big\{t>0: E\exp\big(|U|^\alpha/t^\alpha\big)\leq 2\big\}$. Then, the following technical conditions are necessarily needed. 
\begin{spacing}{1.2}
	\begin{enumerate}[resume,label=\textbf{(C\arabic*)},start=1]\setlength{\itemsep}{-2pt}
		\item\label{con1} \sc{(Sub-Weibull distribution) }\normalfont Assume that both $Z_{ik}$ and $\omega_{ij} $ independently follow sub-Weibull($\alpha$) distributions with $\alpha\in (0,2]$ for every $1\leq k\leq d$ and $1\leq j\leq p$, and are independent of $X_i$. Furthermore, assume that there exists a positive and fixed constant $C_{\text{sw}}$ such that $ \max_k\|Z_{ik}\|_{\psi_\alpha}\leq C_{\text{sw}}$ and $ \max_j\|\omega_{ij}\|_{\psi_\alpha}\leq C_{\text{sw}}$.
		\item\label{con2} \sc{(Loading matrix) }\normalfont Assume that the loading matrix $ B^*=(b_{jk}^*)\in\mR^{p\times d}$ is fixed and there exists a fixed constant $C_B>0$ such that $\max_{j,k}|b_{jk}^*|\leq C_B$.
		\item\label{con3} \sc{(Bounded parameters) }\normalfont Assume that there exist some positive constants $0<\beta_{\min}<C_{\beta\max}<\infty$ and $0< \tau_{\min}\leq\tau_{\max}<1 $ such that (1) $\max_{j}\Vert \beta_{j}^*\Vert \leq C_{\beta\max}$; (2) $\beta_{\min}\leq\min_{j,k\in\mS_{(j),T}}\vert\beta_{jk}^*\vert$; and (3) $ \tau_{\min}^2\leq\min_{j}\tau_{jj}^*\leq \max_{j}\tau_{jj}^*\leq \tau_{\max}^2$.
	\end{enumerate}
\end{spacing}
\noindent The sub-Weibull distribution assumption imposed by Condition \ref{con1} allows for heavier tails and is thus weaker than the popularly used sub-Gaussian assumption in high-dimensional literature \citep{wainwright2019high}. However, it is stronger than the moment conditions widely used in the classical SAR literature \citep{lee2004asymptotic,zhu2020multivariate}. Condition \ref{con1} is necessary in our setting since we are dealing with a problem with a diverging dimension. Consequently, appropriate uniform convergence results are inevitably needed; see for example the uniform consistent result of Theorem \ref{thm1:rho_cmle}. This also explains why similar tail conditions like \ref{con1} have been seldom used in the classical SAR literature of a fixed dimension \citep{lee2004asymptotic,zhu2020multivariate}. However, they are extensively used in the high-dimensional literature \citep{wainwright2019high}. Moreover, Condition \ref{con1} assumes that both $Z_{ik}$ and $\omega_{ij} $ are independent of the exogenous $X_i$. As a consequence, the regression effect term $\mX_{(j)}\beta_{(j)}$ for every $1\leq j\leq p$ can be interpreted in the same way as the usual SAR model \citep{yang2017identification}.

Condition \ref{con2} requires that the true value of the factor loading matrix $B^*$ to be elementwise uniformly bounded \citep{fan2008high,bai2012statistical}. Condition \ref{con3} assumes that: (1) $\Vert \beta_{j}^*\Vert s$ are uniformly upper bounded, (2) the minimum of non-zero $\vert\beta_{jk}^*\vert s$ are uniformly lower bounded away from 0, and (3) the error variance $\tau_{jj}^*s$ are uniformly bounded away from both 0 and 1 as $p\to\infty$. Similar conditions have been used by \cite{fan2011nonconcave} and \cite{wang2012factor}.
\begin{spacing}{1.2}
	\begin{enumerate}[resume,label=\textbf{(C\arabic*)},start=4]\setlength{\itemsep}{-2pt}
		\item\label{con4} \sc{(Diverging response dimension) }\normalfont Assume that (1) $\sqrt{n}/p \to 0$ as $n\to\infty$ and (2) $\log p = O(n^{\alpha\gamma})$, where $\alpha\in (0,2]$ is the sub-Weibull parameter specified in Condition \ref{con1} and $\gamma\in(0,1/4)$ is some fixed constant.
		\item\label{con5} \sc{(Diverging feature dimension) }\normalfont Assume that (1)$(q\log q)^{1/\alpha}/n^{1/2-2\gamma} \to 0$ as $n\to\infty$, and (2) $q/\{(\log q)^{2/\alpha}\log n\}\to 0$ as $n\to\infty$, where $\alpha\in (0,2]$ is the sub-Weibull parameter specified in Condition \ref{con1} and $\gamma\in(0,1/4)$ is defined in Condition \ref{con4}.
\end{enumerate}
\end{spacing}
\noindent The first part of Condition \ref{con4} requires the response dimension $p$ to be sufficiently large so that the latent factors can be estimated consistently \citep{fan2022learning}. The second part of Condition \ref{con4} allows the response dimension $p$ to diverge at an exponentially fast rate \citep{fan2008high,fan2022learning}. By Condition \ref{con5}, we require that the diverging rate of the feature dimension $q$ cannot be too fast \citep{fan2008high,wang2009shrinkageB,cho2013model}.

Next, let $ \big\Vert A\big\Vert_{1} = \max_{1\leq j\leq n}\big\vert \sum_{i=1}^{m}a_{ij}\big\vert$ and $ \big\Vert A\big\Vert_{\infty} = \max_{1\leq i\leq m}\big\vert \sum_{j=1}^{n}a_{ij}\big\vert$ stand for the $\ell_1$-norm (i.e., the maximum absolute column sum) and $\infty$-norm (i.e., the maximum absolute row sum) of an arbitrary matrix $A=\big(a_{ij}\big)\in\mR^{m\times n}$, respectively. In this paper, we use $``\to_d"$ to denote ``convergence in distribution'' and $``\to_p"$ to denote ``convergence in probability''. Write $G_j = WS_j^{-1}$ with $S_j = I_n-\rho_jW$. Define  $\wt{\mX}_{\beta}^*=\big(G_1\mX\beta_1^*,\ldots, G_p\mX\beta_p^*\big)$ $\in\mR^{n\times p} $ and $ \wt{\mZ}_b^* = \big(G_1\mZ b_1^*,\ldots, G_p\mZ b_p^*\big)\in\mR^{n\times p}$. 
\begin{spacing}{1.2}
\begin{enumerate}[resume,label=\textbf{(C\arabic*)},start=6]\setlength{\itemsep}{-2pt}
		\item\label{con6} \sc{(Network matrix) }\normalfont Assume that there exists a sufficiently large but fixed constant $C_W>0$ such that $\big\Vert W\big\Vert_{1}\leq C_W$ and $\max_{j}\big(\big\Vert S_j^{-1}\big\Vert_{1}, \big\Vert S_j^{-1}\big\Vert_{\infty}\big)\leq C_W$ uniformly in $\rho_j\in[-\rho_{\max},\rho_{\max}]$ for some fixed constant $\rho_{\max}\in(0,1)$. 
		\item\label{con7} \sc{(Law of large numbers) }\normalfont Assume that some positive and fixed constants $\kappa_{Gj1}, \kappa_{Gj2}, \kappa_{Gj3}, \kappa_{GGj1}, \kappa_{GGj2}$, and $ \kappa_{Gjd}$, such that (1) $\max_{j}\big| \tr(G_j^{\nu})/n - \kappa_{Gj\nu}\big| = o(1)$ for $\nu =1,2,3$; (2) $\max_{j}\big| \tr\big\{\big(G_j^{\top}G_j\big)^{\nu}\big\}/n - \kappa_{GGj\nu}\big| = o(1)$ for $\nu=1,2$; and (3) $\max_{j}\big| \tr\big\{\diag^2\big(G_j\big)\big\}/n - \kappa_{Gjd}\big| = o(1)$ as $n\to\infty$ uniformly in $\rho_j\in[-\rho_{\max},\rho_{\max}]$ for the same $\rho_{\max}$ given in Condition \ref{con6}. 
		\item\label{con8} \sc{(Identification) }\normalfont Assume a fixed and non-singular matrix $\Sigma_{\mX \mZ}^*\in\mR^{(q+d+2p)\times (q+d+2p)}$ such that (1) $\big\Vert\big(\mX, \wt{\mX}_{\beta}^*, $ $\mZ, \wt{\mZ}_b^*\big)^{\top}\big(\mX, \wt{\mX}_{\beta}^*, \mZ, \wt{\mZ}_b^*\big)/n-\Sigma_{\mX \mZ}^*\big\Vert =o_p(1)$, and (2) $\nu_{\min}\leq\lambda_{\min}\big(\Sigma_{\mX \mZ}^*\big) \leq \lambda_{\max}\big(\Sigma_{\mX \mZ}^*\big)\leq \nu_{\max}$ for some fixed constants $\nu_{\min}>0$ and $\nu_{\max}>0$.
	\end{enumerate}
\end{spacing}
\noindent Condition \ref{con6} assumes that both $W$ and $S_j^{-1}$ are uniformly bounded in both the column and row sums as $n\to\infty$. Condition \ref{con7} is a set of Law of Large Numbers type conditions. Condition \ref{con8} is a sufficient identification condition for $\theta_j^*$ and $b_j^*$ with $1\leq j\leq p$. All those conditions are fairly standard in the literature of spatial autoregression \citep{lee2004asymptotic,lee2010specification,yang2017identification}.

Then, the uniform convergence of $\wh{\theta}_{j,\tcmle}$ is given in Theorem \ref{thm1:rho_cmle}, which is proved in Appendix A.1. By Theorem \ref{thm1:rho_cmle}, we know that $\wh{\theta}_{j,\tcmle}$ is uniformly consistent for $\theta_j^*$ over $1 \leq j \leq p$. The uniform convergence rate is slightly slower than the standard rate of $ 1/\sqrt{n}$ by a factor $(\log p)^{1/\alpha}$. This is the price paid for uniform convergence \citep{fan2013large,fan2022estimating}. In the case of $\alpha=2$ (i.e., sub-Gaussian), this uniform convergence rate becomes $\sqrt{\log p/n}$, which is consistent with the classical results in the existing literature \citep{fan2012variance,wang2012factor,fan2013large}. However, this uniform convergence rate becomes slower if $0<\alpha<2$ with heavier distribution tails.
\begin{theorem}\label{thm1:rho_cmle}
	Assume the conditions \ref{con1}--\ref{con8} hold,  we then have 
\begin{equation*}
	\max_{1\leq j\leq p}\big\Vert\wh{\theta}_{j,\normalfont  \tcmle}-\theta_j^*\big\Vert=O_p\big((\log p)^{1/\alpha}/\sqrt{n} \big).
\end{equation*}
\end{theorem}

\csubsection{Latent Factor Estimation by Diversified Projections}\label{subsec_FactorDP}

Next, consider how to estimate the latent factors in model (\ref{equ:factor}). Following the idea of \cite{fan2022learning}, we develop a method of diversified projections for factor estimation. Specifically, let $M = (m_{jk})\in\mR^{p\times d_{\max}}$ be a pre-specified projection matrix such that $M^{\top}M/p\to \Sigma_M\in\mR^{d_{\max}\times d_{\max}}$ for some positive definite matrix $\Sigma_M$ as $p\to \infty$. Here $d_{\max}\geq d$ is a pre-specified working number of factors. Then, the latent factor $Z_i$ can be estimated by $M^{\top}\varepsilon_i/p= HZ_i + M^{\top}\omega_i/p$ up to an $d_{\max}\times d$ affine transformation $H = M^{\top}B^*/p\in\mR^{d_{\max}\times d}$ and an estimation error $M^{\top}\omega_i/p$. Recall that $\bE=(\varepsilon_{ij})=(\varepsilon_1,\ldots,\varepsilon_n)^{\top}=(\mE_1,\ldots,\mE_p)\in\mR^{n\times p} $, where $\varepsilon_i=(\varepsilon_{ij})\in\mR^p$ is the $i$-th row vector of $\bE$ and $\mE_j=(\varepsilon_{ij})\in\mR^n$ is the $j$-th column vector of $\bE$. Note that $\mE_j = \big(I_n- \rho_j^* W\big)\mY_j - \mX_{(j)}\beta_{(j)}^*$ by model (\ref{model-SAR-new}). Then, a natural estimator for $\mE_j$ can be formed as $\wh{\mE}_j =\big(I_n- \wh{\rho}_{j,\tcmle}W\big)\mY_j - \mX_{(j)}\wh{\beta}_{(j),\tcmle}$. This leads to an estimated residual matrix $\widehat{\mathbb{E}} = (\widehat{\varepsilon}_{ij}) \in \mR^{n\times p}$, which serves an estimator for $\mathbb{E}= (\varepsilon_{ij})\in\mR^{n\times p}$. In practice, $Z_i$ can be estimated by $\wh{Z}_i = M^{\top}\wh{\varepsilon}_i/p$. However, whether the estimation error between $\wh{Z_i}$ and $Z_i$ is asymptotically negligible is not clear. Therefore, we are motivated to study the asymptotic behaviors of $\wh{Z}_i$ rigorously.

To this end, one more technical condition is needed. For an arbitrary matrix, define $\Vert A\Vert =\lambda_{\max}^{1/2}(A^{\top}A) $. Then following \cite{fan2022learning}, we further impose the following technical condition.
\begin{spacing}{1.2}
	\begin{enumerate}[resume,label=\textbf{(C\arabic*)},start=9]\setlength{\itemsep}{-2pt}
		\item\label{confactor} \sc{ }\normalfont Assume that (1) $\max_{1\leq j\leq p}\vert m_{jk}\vert \leq C>0$ for every $1\leq k\leq d_{\max}$ with some positive constant $C$, and (2) $\rank(H) = d$, $\lambda_{\min}(H^{\top}H)\gg 1/p$ and $\lambda_{\max}(H^{\top}H)\leq C\lambda_{\min}(H^{\top}H)$ with $H=M^{\top}B^*/p\in\mR^{d_{\max}\times d}$.
	\end{enumerate}
\end{spacing}
\noindent Condition \ref{confactor} is a combination of Assumption 2.1 and Assumption 2.2 in \cite{fan2022learning}, focusing on the projection matrix (i.e., $M$) and transformation matrix (i.e., $H$). Specifically, the first part of Condition \ref{confactor} requires that the projection matrix $M$ should be uniformly bounded elementwise. The second part of Condition \ref{confactor} prevents $M$ from being orthogonal to $B^*$. Otherwise, the projected random variable $M^{\top}\varepsilon_i/p$ becomes uncorrelated with the latent factor of interest $Z_i$. In that case, we lose the opportunity to estimate $Z_i$ consistently. Write $\wh{\mZ} = \big(\wh{Z}_1,\ldots,\wh{Z}_n\big)^{\top}\in\mR^{n\times d_{\max}}$. Then, we have the following theorem. 
\begin{theorem}\label{thm:factor}
	Assume conditions \ref{con1}--\ref{confactor} hold,  we then have
	$$\Vert\wh{\mZ}-\mZ H^{\top}\Vert/\sqrt{n}=O_p\big(1/\sqrt{n}+1/\sqrt{p}\big).$$ 
\end{theorem}
\noindent The detailed proof of Theorem \ref{thm:factor} is given in Appendix A.2. By Theorem \ref{thm:factor}, we know that $\Vert\wh{\mZ}-\mZ H^{\top}\Vert/\sqrt{n}$ converges to 0 at a rate $O_p\big(1/\sqrt{n}+1/\sqrt{p}\big)$ with $\mZ=\big(Z_1,\ldots,Z_n\big)^{\top}\in\mR^{n\times d}$. This convergence rate contains two parts. The first part $1/\sqrt{n}$ is due to the estimation error of the initial estimator $\wh{\theta}_{j,\tcmle}$ and the second part $1/\sqrt{p}$ is due to the projection error of the diversified projections. 

It is remarkable that Condition \ref{confactor} plays an important role in Theorem \ref{thm:factor}. Therefore, an appropriate specification of $M$ is practically important. In this regard, \cite{fan2022learning} propose four effective solutions. The first solution is called loading characteristics. In this approach, one can construct $M$ by the observed characteristics related to factor loadings. The second solution is called moving window estimation. In this approach, one needs to divide the data into two parts. One can then construct $M$ by the principal component loadings on the first part and then estimate factors by the second part. The third solution is called initial transformation. One can construct $M$ by an appropriate transformation of the initial observation. The fourth solution is called Hadamard projection, which is based on Walsh–Hadamard matrix from a carefully designed statistical experiment. In this work, we implement a random partition method, which is similar to the second solution of \cite{fan2022learning}.

\csubsection{Factor-Augmented Maximum Likelihood Estimators}\label{subsec_FMLE}

By the SAR model (\ref{model-SAR-new}) and the factor model (\ref{equ:factor}), we obtain a factor-augmented spatial autoregressive (FSAR) model as
\begin{equation}\label{model-FSAR}
	\mathbb{Y}_j = \rho_j W \mathbb{Y}_j + \mX_{(j)}\beta_{(j)}+ \wt{\mZ} \wt{b}_j + \Omega_j,
\end{equation}
where $\wt{\mZ} = \mZ H^{\top}\in\mR^{n\times d_{\max}}$ is the common factor after $H$-transformation, $\wt{b}_j = H\big(H^{\top}H\big)^{-1}$ $b_j\in\mR^{d_{\max}}$, $b_j = (b_{jk})\in\mR^d$ is the $j$-th row vector of $B$, and $\Omega_j=(\omega_{ij})\in\mR^n$ is the independent random noise with mean $0$ and covariance $\tau_{jj} I_n$. For a given $j$, model (\ref{model-FSAR}) is similar to the spatial autoregressive model with additional $X$-covariates (i.e., $\wt{\mZ}$). However, there is a critical difference. That is an ``additional $X$-covariates" here (i.e., $\wt{\mZ}$) is a latent random matrix and cannot be directly observed. A natural solution is to replace $\wt{\mZ}$ by its estimator $\wh{\mZ}$. Then a factor-augmented log-likelihood function can be specified out as
\begin{equation*}
	\mL_{\tfmle}^{(j)}\big(\Theta_j,\wh{\mZ}\big) = -\frac{n}{2}\log \tau_{jj}+\log\big\vert S_j\big\vert- \frac{1}{2\tau_{jj}}\Big(S_j\mY_j- \mX_{(j)}\beta_{(j)}-\wh{\mZ}\wt{b}_j\Big)^\top \Big(S_j\mY_j- \mX_{(j)}\beta_{(j)}-\wh{\mZ}\wt{b}_j\Big),
\end{equation*}
where $\Theta_j = \big(\rho_j, \beta_{(j)}^{\top}, \wt{b}_j^{\top}, \tau_{jj}\big)^{\top}\in\mathbb{R}^{g_j}$ with $g_j=s_j+d_{\max}+2$. By maximizing $\mL_{\tfmle}^{(j)}\big(\Theta_j,\wh{\mZ}\big)$ with respect to $\Theta_j$, we obtain $\wh{\Theta}_{j,\tfmle} = \big(\wh{\rho}_{j,\tfmle}, \wh{\beta}_{(j),\tfmle}, \wh{b}_{j,\tfmle}, \wh{\tau}_{jj,\tfmle}\big)^{\top} = \argmax_{\Theta}$ $\mL_{\tfmle}^{(j)}\big(\Theta,\wh{\mZ}\big) $. Here we refer to $\wh{\Theta}_{j,\tfmle} $ as a factor-augmented maximum likelihood estimator (FMLE). Numerically, the FMLE $\wh{\Theta}_{j,\tfmle}$ with different $j$ can be computed in a fully parallel or distributed way. 

We next consider how to establish the asymptotic properties of $\wh{\Theta}_{j,\tfmle}$. Note that the resulting estimator $\wh{\Theta}_{j,\tfmle}$ is defined as the maximizer of $\mL_{\tfmle}^{(j)}\big(\Theta_j,\wh{\mZ}\big)$ with the estimated latent factors $\wh{\mZ}$. It is unclear whether the estimation error of $\wh{\mZ}$ affects the asymptotic behaviors of $\wh{\Theta}_{j,\tfmle}$. Recall that $\wh{\mZ}$ is a consistent estimator for $\wt{\mZ}$ by Theorem \ref{thm:factor}. Let $\Theta_j^* =  \big(\rho_j^*, \beta_{(j)}^{*\top}, \wt{b}_j^{*\top}, \tau_{jj}^*\big)^{\top}$ denotes the true value of $\Theta_j$ with $\wt{b}_j^{*}=H(H^{\top}H)^{-1}b_j^*$. Then we can apply the Taylor's expansion and obtain the following asymptotic approximation for $\wh{\Theta}_{j,\tfmle}$ as 
\begin{gather}
	\sqrt{n}\Big(\wh{\Theta}_{j,\tfmle}-\Theta_j^*\Big)=\Big\{-\ddot{\mL}_{\Theta_j^*\Theta_j^*}\big(\Theta_j^*,\wt{\mZ}\big)/n\Big\}^{-1}\nonumber\\
	\label{equ:ThetaZ}\Big\{\dot{\mL}_{\Theta_j^*}\big(\Theta_j^*,\wt{\mZ}\big)/\sqrt{n}+\sum_{i=1}^{n}\ddot{\mL}_{\Theta_j^*\wt{Z}_i}\big(\Theta_j^*,\wt{\mZ}\big)\big(\wh{Z}_i - \wt{Z}_i\big)/\sqrt{n} + o_p(1) \Big\},
\end{gather}
where $\dot{\mL}_{\Theta_j}\big(\Theta_{j},\wt{\mZ}\big)=\partial \mL_{\tfmle}^{(j)}\big(\Theta_j,\wt{\mZ}\big)/\partial \Theta_j\in\mR^{g_j}$ and $\ddot{\mL}_{\Theta_j\Theta_j}\big(\Theta_{j},\wt{\mZ}\big)=\partial^2 \mL_{\tfmle}^{(j)}\big(\Theta_j,\wt{\mZ}\big)/\partial \Theta_j$ $\partial\Theta_j^{\top}\in\mR^{g_j\times g_j}$ are the 1st and 2nd order partial derivatives of $ \mL_{\tfmle}^{(j)}\big(\Theta_j,\wt{\mZ}\big)$ with respect to $\Theta_j$, respectively. Here $\ddot{\mL}_{\Theta_j\wt{Z}_i}\big(\Theta_j,\wt{\mZ}\big)= \partial^2 \mL_{\tfmle}^{(j)}\big(\Theta_j,\wt{\mZ}\big)/\partial \Theta_j\partial\wt{Z}_i^{\top}\in\mR^{g_j\times d_{\max}}$ is the 2nd order partial derivative of $ \mL_{\tfmle}\big(\Theta_j,\wt{\mZ}\big)$ with respect to $\Theta_j$ and $\wt{Z}_i$. Compared with the classical approximation theory \citep{lee2004asymptotic,lee2010estimation}, there involves an extra term $\sum_{i=1}^{n}\ddot{\mL}_{\Theta_j^*\wt{Z}_i}\big(\Theta_j^*,\wt{\mZ}\big)\big(\wh{Z}_i - \wt{Z}_i\big)/\sqrt{n} $ in (\ref{equ:ThetaZ}) due to the estimation of $\wt{\mZ}$. This motivates us to study this extra term rigorously. It can be theoretically verified that 
\begin{gather}
	\dfrac{1}{n}\sum_{i=1}^{n}\ddot{\mL}_{\Theta_j^*\wt{Z}_i}\big(\Theta_j^*,\wt{\mZ}\big)\sqrt{n}\big(\wh{Z}_i - \wt{Z}_i\big) \nonumber\\
	\label{equ:QZZ}= -\dfrac{1}{p\tau_{jj}}\sum_{k=1}^{p}\Big\{c_{jk}^{\Theta*}\sqrt{n}\big(\rho_k^* - \wh{\rho}_{k,\tcmle}\big) + Q_{jk}^{\Theta*}\sqrt{n}\big(\beta_{(k)}^*-\wh{\beta}_{{(k)},\tcmle}\big)\Big \}+o_p(1),
\end{gather}
where $ c_{jk}^{\Theta*} \in\mR^{g_j}$ and $Q_{jk}^{\Theta*}\in\mR^{g_j\times s_k} $ are some unknown parameters defined in (A.23) and (A.24) of Appendix A.3, respectively. 

Note that $ \wh{\rho}_{k,\tcmle}$ and $ \wh{\beta}_{(k),\tcmle}$ for every $1\leq k\leq p$ are the initial estimators defined in Section~\ref{subsec_CMLE}. By (\ref{equ:QZZ}), we find that the estimation error of $\wh{\mZ}$ should play an important role in determining the asymptotic distribution of $\wh{\Theta}_{j,\tfmle}$ through $ \wh{\rho}_{k,\tcmle}$ and $ \wh{\beta}_{(k),\tcmle}$. Then, the asymptotic properties of $\wh{\Theta}_{j,\tfmle}$ can be well studied by combining (\ref{equ:ThetaZ}) and (\ref{equ:QZZ}). To this end, one more technical condition is needed.
\begin{spacing}{1.2}
	\begin{enumerate}[resume,label=\textbf{(C\arabic*)},start=10]\setlength{\itemsep}{-2pt}
		\item\label{con10} \sc{ }\normalfont Assume that $\sum_{k=1}^p |\tau_{jk}^*| = O(1) $, $\lambda_{\max}\big(\mA_j/p^2\big) = o(1)$, and $\lambda_{\min}\big(\mD_j/$ $p^2\big) \geq C_d$ for every $1\leq j\leq p$, where $C_d>0$ is a fixed constant, $ \mA_j$ and $ \mD_j$ are some matrices defined in (A.30) and (A.33) of Appendix A.3, respectively. 
	\end{enumerate}
\end{spacing}
\noindent Condition \ref{con10} puts one particular type of sparsity constraint on the covariance matrix $\Sigma_{\omega}^*=\big(\tau_{jk}^*\big)\in\mR^{p\times p}$ \citep{fan2008high,bai2012statistical,wang2012factor}. It can be well satisfied for many important special cases, such as $\Sigma_{\omega} = \diag\big(\tau_{11},\ldots,\tau_{pp}\big)\in\mR^{p\times p}$. We then have the following theorem about the asymptotic behavior of $\wh{\Theta}_{j,\tfmle}$, which is proved in Appendix A.3.
\begin{theorem}\label{thm:Theta}
	Assume conditions \ref{con1}--\ref{con10} hold, we then have  $\sqrt{n}\big(\wh{\Theta}_{j,\tfmle}$ $ -\Theta_j^*\big)\to_d N\big(0, \Sigma_{2\Theta_j^*}^{-1}\Sigma_{1\Theta_j^*}^*\Sigma_{2\Theta_j^*}^{-1}\big)$ as $n\to\infty$, where $ \Sigma_{1\Theta_j^*}^* =\Sigma_{2\Theta_j^*} +\Delta_{\Theta_j^*}+ \Sigma_{\mQ_j}\in\mR^{g_j\times g_j}$, 
\begin{equation*}
	\renewcommand{\arraystretch}{0.8}
	\setlength{\arraycolsep}{2pt}
	\Sigma_{2\Theta_j^*} =
\begin{pmatrix}
	\Sigma_{2\rho_j^*\rho_j^*}&\Sigma_{2\beta_{(j)}^*\rho_j^*}^{\top}&\Sigma_{2\wt{b}_j^*\rho_j^*}^{\top}&\Sigma_{2\tau_{jj}^*\rho_j^*}\\
	\Sigma_{2\beta_{(j)}^*\rho_j^*} &\Sigma_{2\beta_{(j)}^*\beta_{(j)}^*}&0_{q,d_{\max}}&0_{q}\\
	\Sigma_{2\wt{b}_j^*\rho_j^*} &0_{d_{\max},q}&\Sigma_{2\wt{b}_j^*\wt{b}_j^*}&0_{d_{\max}}\\
	\Sigma_{2\tau_{jj}^*\rho_j^*}&0_{q}^{\top}&0_{d_{\max}}^{\top}&\Sigma_{2\tau_{jj}^*\tau_{jj}^*}\\
\end{pmatrix},
\Delta_{\Theta_j}=
\begin{pmatrix}
	\Delta_{\rho_j^*\rho_j^*}&\Delta_{\beta_{(j)}^*\rho_j^*}^{\top}&\Delta_{\wt{b}_j\rho_j^*}^{\top}&\Delta_{\tau_{jj}^*\rho_j^*}\\
	\Delta_{\beta_{(j)}^*\rho_j^*}& 0_{q,q}&0_{q,d_{\max}}&\Delta_{\tau_{jj}^*\beta_{(j)}^*}\\
	\Delta_{\wt{b}_j^*\rho_j^*} &0_{d_{\max},q} & 0_{d_{\max},d_{\max}} & \Delta_{\tau_{jj}^*\beta_{(j)}^*}\\
	\Delta_{\tau_{jj}^*\rho_j^*} & \Delta_{\tau_{jj}^*\beta_{(j)}^*}^{\top}& \Delta_{\tau_{jj}^*\wt{b}_j^*}^{\top}&  \Delta_{\tau_{jj}^*\tau_{jj}^*}\\
\end{pmatrix},
\end{equation*}
and $\Sigma_{\mQ_j}\in\mR^{g_j\times g_j} $. The analytical expressions of the matrices $\Sigma_{2\Theta_j^*}$, $\Delta_{\Theta_j^*}$, and $\Sigma_{\mQ_j} $ are given in Appendix A.3 and Appendix C.3, respectively.
\end{theorem}

By Theorem \ref{thm:Theta}, we know that $\wh{\Theta}_{j,\tfmle}$ is $\sqrt{n}$-consistent and asymptotically normal. Note that asymptotic covariance of $\wh{\Theta}_{j,\tfmle}$ consists of three parts. The first part $ \Sigma_{2\Theta_j^*} $ represents a typical information matrix under normality. The second part $ \Delta_{\Theta_j^*}$ contains high order moments of the disturbances \citep{lee2004asymptotic,lee2010estimation,yang2017identification}. This part becomes zero if $\Omega_j$ in (\ref{model-FSAR}) follows a normal distribution strictly. Moreover, the last term $\Sigma_{\mQ_j}$ is due to the estimation error of $\wh{\mZ}$. We should have $ \Sigma_{\mQ_j} =0$ if the true $\wt{\mZ}$ were actually observed. In this case, $\wh{\Theta}_{j,\tfmle}$ becomes statistically as efficient as the oracle estimator $\wt{\Theta}_{j,\tfmle} = \argmax_{\Theta}\mL_{\tfmle}^{(j)}\big(\Theta,\wt{\mZ}\big)$.

\csubsection{Shrinkage Estimation and Uniform Selection Consistency}\label{subsec_Shrink}

Next, we consider how to consistently estimate the unknown $\mS_{(j),T}$ for every response $\mY_j$ in practice. To this end, various shrinkage estimation techniques can be considered \citep{tibshirani1996regression,fan2001variable,fan2021shrinkage}. In this work, we focus on the smoothly clipped absolute deviation (SCAD) method of \cite{fan2001variable} and \cite{fan2011nonconcave} due to its excellent theoretical properties. The methodology developed below can be readily applied to other popular shrinkage methods without additional difficulty \citep{tibshirani1996regression,efron2004least,fan2021shrinkage}. Specifically, we define a penalized likelihood function for the model (\ref{model-SAR}) as $\mQ_{\lambda}^{(j)}\big(\theta_j\big) = \mL^{(j)}\big(\theta_j\big) -n\sum_{k=1}^{q}p_{\lambda}\big(|\beta_{jk}|\big)$, where $ \mL^{(j)}\big(\theta_j\big) = -n \big(\log \sigma_{jj}\big)/2 + \log\big\vert S_j\big\vert- \big(S_j \mY_j - \mX\beta_j \big)^{\top}\big(S_j \mY_j - \mX\beta_j \big)/\big(2\sigma_{jj}\big)$ and $p_{\lambda}(\cdot) $ is the SCAD penalty function with its first order derivative given by $ \dot{p}_{\lambda}\big(t\big) = \lambda\big[I(t\leq \lambda) +  \big(a\lambda-t\big)_{+}I(t>\lambda)/\big\{(a-1)\lambda\big\}\big]$. Here $a$ is some constant that is often taken to be 3.7 \citep{fan2001variable,fan2011nonconcave}, $\lambda$ is a tuning parameter, $(t)_{+} = tI(t>0)$, and $I(\cdot)$ is the indicator function. Then,  a SCAD estimator can be obtained as $ \wh{\theta}_{j,\lambda} =\big(\wh{\rho}_{j,\lambda}, \wh{\beta}_{j,\lambda}^{\top}, \wh{\sigma}_{j,\lambda}\big)^{\top}= \argmax_{\theta}\mQ_{\lambda}^{(j)}\big(\theta\big)$.

As demonstrated by \cite{fan2001variable} and many subsequent works \citep{fan2011nonconcave,fan2020statistical}, the SCAD estimator has excellent model selection capabilities for various statistical models. It is then of interest to study whether similar properties can be reproduced in our case. Following the literature \citep{wang2007tuning,wang2009shrinkageB}, write $\lambda_n$ as a tuning parameter sequence indexed by $n$. Accordingly, define $\wh{\mS}_{(j),\lambda_n} = \big\{1\leq k\leq q:  \wh{\beta}_{jk,\lambda_n}\neq 0\big\}$ as the model set selected by $ \wh{\theta}_{j,\lambda_n} $. Recall that $\beta_{(j)} = \big\{ \beta_{jk}: k\in \mS_{(j),T} \big\} \in\mR^{s_j}$ is the sub-vector associated with the nonzero coefficients. Define $\beta_{(-j)} = \big\{ \beta_{jk}: k\notin \mS_{(j),T} \big\} \in\mR^{q-s_j}$ to be the sub-vector associated with the zero coefficients for every $1\leq j\leq p$. Write $ \wh{\beta}_{j,\lambda_n}= \big(\wh{\beta}_{(j),\lambda_n}^{\top}, \wh{\beta}_{(-j),\lambda_n}^{\top}\big)^{\top} \in\mR^q$ and $\beta_{j}^* = \big(\beta_{(j)}^{*\top}, \beta_{(-j)}^{*\top}\big)^{\top} \in\mR^q$ with $ \beta_{(j)}^* \neq 0 $ and $\beta_{(-j)}^* = 0$. Then, the following theorem establishes the uniform selection consistency of $ \wh{\mS}_{(j),\lambda_n}$ over $1\leq j\leq p$. 
\begin{theorem}\label{thm4}
	Assume the conditions \ref{con1}--\ref{con10} hold. Further assume that $\lambda_n\to 0$ and $\sqrt{n}\lambda_n/\big\{\log(pq)^{1/\alpha}\big\}\to \infty$ as $n\to\infty$, we then have 
	\begin{equation*}
		P\Big(\wh{\mS}_{(j),\lambda_n} = \mS_{(j),T},\text{\normalfont for every } 1\leq j\leq p\Big) \to 1.
	\end{equation*}
\end{theorem} 

\noindent The detailed proof of Theorem \ref{thm4} is provided in Appendix A.4. By Theorem \ref{thm4}, we know that, with probability tending to one, the selected set $\wh{\mS}_{(j),\lambda_n}$ consistently recovers the true set $\mS_{(j),T}$ exactly in a way uniformly over $1\leq j\leq p$. It is remarkable that this uniform selection consistency result is stronger than the conventional selection consistency discussed in the existing literature \citep{shao1993linear,wang2007tuning,wang2009shrinkageB}, which focuses on a single $j\in\{1,\ldots,p\}$.

We next consider how to specify $\lambda_n$ practically. To this end, a number of Bayesian information criterion (BIC) methods have been developed under various model setups \citep{wang2007tuning,chen2008extended,wang2009shrinkageB,wang2012factor}. We are then inspired to develop for our model (\ref{model-SAR}) a similar BIC-type criterion as 
\begin{equation*}
	\text{BIC}^{(j)}(\lambda) = -\dfrac{1}{n}\mL^{(j)}\big(\wh{\theta}_{j,\lambda}\big) +\dfrac{1}{n} \big|\wh{\mS}_{(j),\lambda}\big|\big(\log n\big)\Big\{\log(pq)\Big\}^{2/\alpha}.
\end{equation*}
Note that this BIC criterion contains two components. The first component $ \mL^{(j)}\big(\wh{\theta}_{j,\lambda}\big)/n$ reflects the goodness-of-fit. The second component penalizes the model complexity $\big|\wh{\mS}_{(j),\lambda}\big| $ by a factor $ (\log n)\big\{\log(pq)\big\}^{2/\alpha}/n$. The first factor $\log n$ is due to the diverging sample size $n$, and the second factor $\big\{\log(pq)\big\}^{2/\alpha}/n $ is due to the diverging feature $q$ and the diverging response $p$. Penalizing factors of a similar form have been popularly used in the literature \citep{chen2008extended,wang2009shrinkageB,zhang2024variable}.

Then, an optimal tuning parameter can be selected as $\wh{\lambda}_{(j),\text{BIC}} = \argmin_{\lambda}\text{BIC}^{(j)}(\lambda) $. Note that $\wh{\lambda}_{(j),\text{BIC}} $ is an estimator depending on $n$. This leads to a selected model as $ \wh{\mS}_{(j),\wh{\lambda}_{(j),\text{BIC}}} $. To study its uniform selection consistency property, define $\mS_F = \big\{1,\ldots,q\big\}$ as the full model. Write $\mS_{(j)} \subset \mS_F$ with size $q_j=\big\vert \mS_{(j)}\big\vert $ as an arbitrary working model for the $j$-th response. Define $\mathcal{R}_{(j)}(\theta) = E\big\{-\mL^{(j)}(\theta)/n\big\}$ as the risk function, and let $\mathcal{R}_{(j),\min}^* = \mathcal{R}_{(j)}\big(\theta_{j}^*\big)$ be its minimum value evaluated at the true parameter. Then, the following technical condition is necessarily needed.
\begin{spacing}{1.2}
	\begin{enumerate}[resume,label=\textbf{(C\arabic*)},start=11]\setlength{\itemsep}{-2pt}
		\item\label{con11} \sc{ }\normalfont Assume that there exists some positive and fixed constant $\delta_{\min}>0$ such that $\min_{1\leq j\leq p}\min_{\mS_{(j)}\not\supset\mS_{(j),T}}\inf_{\theta_{j,\mS_{(j)}}}\big\{\mathcal{R}_{(j)}\big(\theta_{j,\mS_{(j)}}\big) - \mathcal{R}_{(j),\min}^* \big\}\geq \delta_{\min}$.
	\end{enumerate}
\end{spacing}
\noindent Condition \ref{con11} imposes a strict separation condition on the risk function $\mathcal{R}_{(j)}(\theta) $. It ensures that the minimal risk of any underfitted working model (i.e., $\mS_{(j)}\not\supset\mS_{(j),T}$) must be strictly larger than that of the true model by a fixed margin $\delta_{\min}$. Similar conditions have been widely used in the literature; see for example Condition (2.5) in \cite{shao1993linear}, Condition 2 in \cite{wang2007tuning}, and Assumption 2 in \cite{fan2012variance}. Then, the uniform selection consistency of $\mS_{(j),\wh{\lambda}_{(j),\text{BIC}}} $ can be rigorously established by Theorem \ref{thm5}, whose detailed proof is given in Appendix A.5. By Theorem~\ref{thm5}, we know that, with probability tending to one, the selected model $ \wh{\mS}_{(j),\wh{\lambda}_{(j),\text{BIC}}} $ recovers the true model $\mS_{(j),T}$ uniformly over $1\leq j\leq p$. 
\begin{theorem}\label{thm5}
	Assume the conditions \ref{con1}--\ref{con11} hold, we then have as $ n\to\infty$, 
	\begin{equation*}
		P\Big(\wh{\mS}_{(j),\wh{\lambda}_{(j),\text{\normalfont BIC}}}  = \mS_{(j),T},\text{\normalfont for every } 1\leq j\leq p\Big) \to 1.
	\end{equation*}
\end{theorem} 

As we mentioned before, these true model sets $\mathcal{S}_{j,T}s$ are practically unknown and therefore have to be empirically estimated by $\wh{\mS}_{(j),\wh{\lambda}_{(j),\text{\normalfont BIC}}}s$. Once they are empirically estimated, they are then treated as if they were the truth. Thereafter, the three-step estimators as developed in Sections \ref{subsec_CMLE}--\ref{subsec_FMLE} can be readily computed. This leads to the final empirical estimators. Strictly speaking, the empirical estimators finally computed are different from the oracle estimators studied in Sections \ref{subsec_CMLE}--\ref{subsec_FMLE}, since there exists a positive probability for $\wh{\mS}_{(j),\wh{\lambda}_{(j),\text{\normalfont BIC}}}\neq\mathcal{S}_{j,T}$. Nevertheless, this probability shrinks to zero as $n\to\infty$ due to the uniform selection consistency results as established in Theorems \ref{thm4} and \ref{thm5}. Therefore, the two estimators (i.e., the oracle estimators and the empirical estimators) share the same asymptotic distribution. Therefore, we are able to claim that both estimators are equivalent asymptotically.

\csection{NUMERICAL STUDIES}\label{sec_numer}

\csubsection{Simulation Models}\label{subsec_simulation}

To demonstrate the finite sample performance of the FSAR model, we conduct a number of simulation studies. For each simulation replication, we first generate the adjacency matrix $A=(a_{i_1i_2})\in\mR^{n\times n}$, and then set the diagonal element $a_{ii} = 0$ for every $1\leq i\leq n$. Note that $A$ is not necessarily a symmetric matrix. Thereafter, the adjacency matrix $A$ is row-normalized as $w_{i_1i_2} = a_{i_1i_2}/n_{i_1}$ for each row $1\leq i_1\leq n$. This leads to the spatial weight matrix $W=(w_{i_1i_2})\in\mR^{n\times n}$. Regarding the adjacency matrix $A$, three widely standard network structures are considered. 

\sc{Example 1.}\normalfont$\ $(Dyad Independence Model, DIM) Following \cite{holland1981exponential}, define a dyad as $\mA_{i_1i_2} = (a_{i_1i_2}, a_{i_2i_1})$ for any $1\leq i_1<i_2\leq n$. Different $\mA_{i_1i_2}s$ are assumed to be mutually independent. Next, following \cite{zhu2020multivariate}, define $P\big\{\mA_{i_1i_2}=(1,1)\big\}=2n^{-1}$ and $P\big\{\mA_{i_1i_2}=(1,0)\big\}=P\big\{\mA_{i_1i_2}=(0,1)\big\}=0.5n^{-0.8}$. As a result, the expected number of the mutually connected dyads with $\mA_{i_1i_2}=(1,1)$ is $O(n)$. In the meanwhile, the expected degree of each node to be slowly diverging in the order of $O(n^{0.2})$. Then, we have $P\big\{\mA_{i_1i_2}=(0,0)\big\}=1-2n^{-1}-n^{-0.8}$, which is close to 1 as the network size $n\to\infty$.

\sc{Example 2.}\normalfont$\ $(Stochastic Block Model, SBM) Consider a network structure generated from the stochastic block model. Specifically, set $K = 5$ be the total number of blocks. Next, following \cite{nowicki2001estimation}, we randomly assign each node a block label ($k = 1,\ldots, K$) with equal probability $1/K$. Let $P(a_{i_1i_2}=1)=9/n$ if $i_1$ and $i_2$ belongs to the same block, and $P(a_{i_1i_2}=1)=3/n$ otherwise. Therefore, nodes within the same block are more likely to be connected with each other.

\sc{Example 3.}\normalfont$\ $(Latent Space Model, LSM) Following \cite{hoff2002latent}, assume that the node $i$ has a low-dimensional position $d_i$ in the latent space for every $1\leq i\leq n$. The probability of two nodes being connected (i.e., $P(a_{i_1i_2}=1)$) is determined by the distance between their respective latent positions (i.e., $ d_{i_1i_2} = \Vert d_{i_1}-d_{i_2}\Vert$). Here, we set $P(a_{i_1i_2}=1) =\exp\big(-0.25nd_{i_1i_2}\big)\big/\big\{1+\exp\big(-0.25nd_{i_1i_2}\big)\big\} $, where $d_i$ is independently and identically uniformly distributed on $(0,1)$ for every $1\leq i\leq n$.

For each network structure, consider multiple network sizes (i.e., $n=500$, 1000 and 1500), response dimensions (i.e., $p=50$, 100, and 200), covariate dimensions (i.e., $q=5$, 10, and 20), and latent factor dimensions (i.e., $d=1$, 2, and 3). Next, generate $\omega_i=(\omega_{i1},\ldots,\omega_{ip})^{\top}\in\mR^p$ with $\omega_{ij}$ independently drawn from $N(0,\tau_{jj})$, where $\tau_{jj}$ is independently generated from a uniform distribution $U(0.1,0.2)$. Then, both $X_i\in\mR^q$ and $Z_i\in\mR^{d}$ are sampled from a standard multivariate normal distribution. The true model size of each response is set to be $s_j = 2$ for $q=5$, $s_j= 5$ for $q=10$, and $s_j = 10$ for $q=20$. Then, for $\mB = (\beta_{jh})\in\mR^{p\times q}$, we independently sample $\beta_{jh}s$ from $U(0.5,1)$ if $1\leq h\leq s_j$, and $0$ otherwise. Then, for the true parameters $B=(b_{jk})\in\mR^{p\times d}$ and $\rho=(\rho_1,\ldots,\rho_p)^{\top}\in\mR^p$, we independently sample $b_{jk}s$ from $N(0,1)$, and $\rho_js$ from $U(0.2,0.9)$, respectively. Accordingly, the high-dimensional response matrix $Y\in\mR^{n\times p}$ can be obtained according to model (\ref{model-FSAR}).

\csubsection{Simulation Results}\label{subsec_simu_re}

\begin{figure}[tb]
	\centering
	\includegraphics[width=6in]{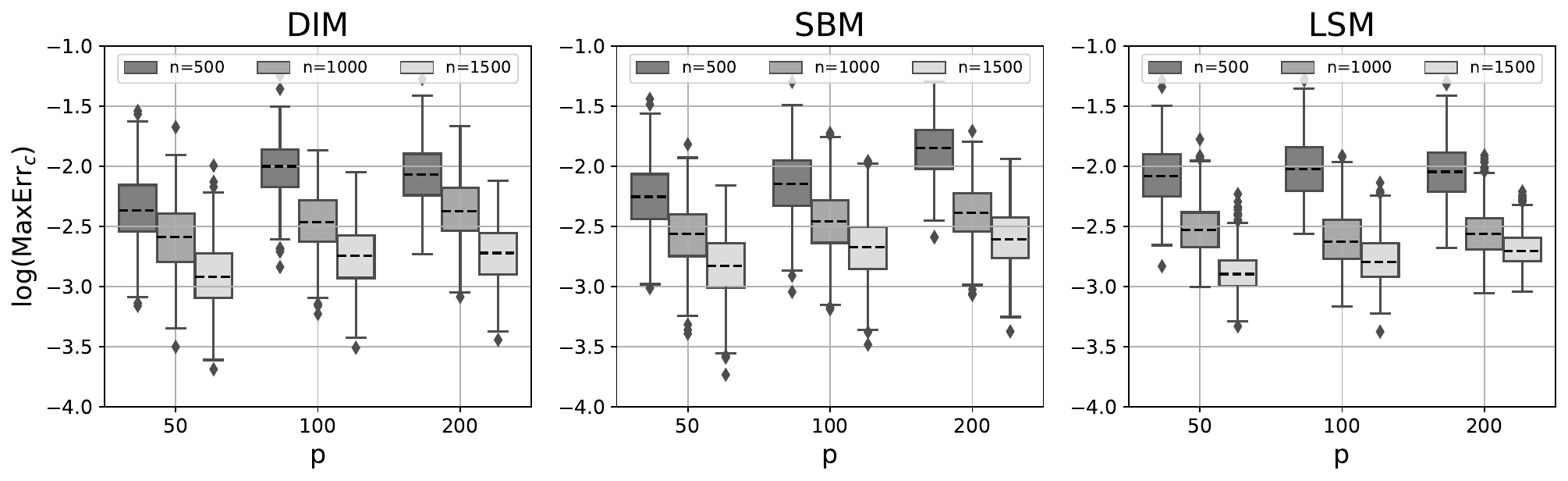}
	\caption{The log(MaxErr$_c$) values for CMLE $\wh{\rho}_{j,\tcmle}$ with $d=3$. Different panels correspond to different network structures:  DIM (the left), SBM (the middle), and LSM (the right). For a given panel, different groups correspond to different feature dimensions with $p=50$, $100$ and $200$, respectively. For a given group, the lighter the color of the box is, the larger the sample size is.}
	\label{fig:cmle}
\end{figure}

We start with the uniform convergence of CMLE $\wh{\rho}_{j,\tcmle}s$. Given a specification $(W,n,p,q,d)$, we randomly replicate the experiment for a total of $R=500$ times. Since the simulation results are qualitatively similar, we only report here the case with $(q,d) = (20,3)$. We use $\wh{\rho}_{j,\tcmle}^{(r)}$ to represent one particular estimator obtained in the $r$-th replication ($1\leq r\leq R$). The true parameter is denoted by $\rho_j$. Define the estimation error (Err) as Err$_{j,c}^{(r)}=\vert \wh{\rho}_{j,\tcmle}^{(r)}-\rho_j\vert$ for every $\wh{\rho}_{j,\tcmle}^{(r)}$, and the maximum error (MaxErr) over $j$ as MaxErr$_c^{(r)}=\max_{1\leq j\leq p}$Err$_{j,c}^{(r)}$. This leads to a total of $R$ MaxErr$_c$ values, which are then log-transformed and box-plotted in Figure \ref{fig:cmle}. By Figure \ref{fig:cmle}, we obtain the following two findings. First, for a fixed $W$ and $p$, the maximum error (MaxErr) decreases as the sample size $n$ increases. This provides empirical evidence for the uniform consistency of $\wh{\rho}_{j,\tcmle}$ over $1\leq j\leq p$. Moreover, with a fixed $W$ and $n$ but diverging $p$, the maximum error (MaxErr) increases slowly. This suggests that the uniform convergence rate of $\wh{\rho}_{j,\tcmle}$ diverges with respect to $p$ but at a slow rate. All of these results are in line with our theoretical finding in Theorem \ref{thm1:rho_cmle}. 

We next study the finite sample performance of the FMLE $ \wh{\rho}_{j,\tfmle}s$. To this end, we need to specify the projection matrix $M$. Similar to the moving window estimation method of \cite{fan2022learning}, we implement here a random partition method, which uses 10\% of the randomly generated sample to estimate the projection matrix $M$. Once $M$ is specified, the rest 90\% samples are then used to conduct the subsequent analysis. First, we compute for each $\wh{\rho}_{j,\tfmle}$ an Err value decoded by Err$_{j,f}^{(r)}$ at the $r$-th replication. Then, we obtain the mean error as MErr$_f=(Rp)^{-1}\sum_{r=1}^{R}\sum_{j=1}^{p}$Err$_{j,f}^{(r)}$. For comparison purposes, the same value is also computed for the CMLE and is denoted as MErr$_c$. Next, we compare their relative efficiency by the relative improvement margin RIM = $\big(1-$MErr$_f$/MErr$_c\big)\times 100\%$. Moreover, for each $1\leq j\leq p$, a 95\% confidence interval is constructed for $\rho_j$ as CI$_j^{(r)} = \big(\wh{\rho}_{j,\tfmle} - z_{0.975}\wh{\text{SE}}_j^{(r)}, \wh{\rho}_{j,\tfmle} + z_{0.975}\wh{\text{SE}}_j^{(r)} \big)$, where $ \wh{\text{SE}}_j^{(r)}$ is square root of the first $(1,1)$ component of $\wh{\Sigma}_{2\Theta_j^*}^{-1}\wh{\Sigma}_{1\Theta_j^*}^*\wh{\Sigma}_{2\Theta_j^*}^{-1}$, and $\wh{\Sigma}_{2\Theta_j^*}^{-1}\wh{\Sigma}_{1\Theta_j^*}^*\wh{\Sigma}_{2\Theta_j^*}^{-1}$ is a plug-in estimator of the asymptotic covariance matrix $\Sigma_{2\Theta_j^*}^{-1}\Sigma_{1\Theta_j^*}^*\Sigma_{2\Theta_j^*}^{-1}$ given in Theorem \ref{thm:Theta}. Here $z_{\alpha}$ is the $\alpha$th quantile of a standard normal distribution. Then the coverage probability (CP) is computed as CP$_j = R^{-1}\sum_{r=1}^RI\big(\rho_j\in \text{CI}_j^{(r)}\big)$. Different RIM values for different combinations $(n, p,W)$ are computed and reported in Table \ref{table:large p}. For ease of presentation, a CP$_j$ is randomly selected over $1\leq j\leq p$ and also reported in Table \ref{table:large p}. By Table \ref{table:large p}, we find that given $p$ and $W$, larger sample sizes always lead to smaller MErr$_f$ values. This confirms the consistency of FMLE, which in line with the result in Theorem \ref{thm:Theta}. Compared with CMLE, the estimation efficiency of FMLE is improved by about 35\% on average. Moreover, the reported coverage probability values are all fairly close to the nominal level of 95\%. This implies that the estimated standard error approximates the true standard error very well. Those results provide numerical evidence of the asymptotic theory obtained in Theorem \ref{thm:Theta}.
\begin{table}[tb]
	\large
	\caption{The simulation results of FMLE for three networks.}
	\vspace{0.5em}
	\centering
	\scalebox{0.9}{	
		\begin{tabular}{c|ccc|ccc|ccc}
			\toprule
			\hline
			\multicolumn{1}{c|}{ } & \multicolumn{3}{c|}{$p=50$} & \multicolumn{3}{c|}{$p=100$} &\multicolumn{3}{c}{$p=200$}\\ 
			 \multicolumn{1}{c|}{$n$} & {\small MErr$_f$}& {\small RIM(\%)}&  \multicolumn{1}{c|}{{\small CP(\%)}} & {\small MErr$_f$}& {\small RIM(\%)}& \multicolumn{1}{c|}{{\small CP(\%)}} & {\small MErr$_f$} & {\small RIM(\%)}& \multicolumn{1}{c}{{\small CP(\%)}}
			 \\ \hline 
			 \multicolumn{1}{c|}{ }& \multicolumn{3}{c|}{ } & \multicolumn{3}{c|}{{\small \textbf{DIM}}}&\multicolumn{3}{c}{ } \\
			 \multicolumn{1}{c|}{500}  & 0.018 & 41.58 & \multicolumn{1}{c|}{95.52} & 0.019 &51.98& \multicolumn{1}{c|}{95.40} & 0.020 & 42.75 & \multicolumn{1}{c}{95.20}\\
			 \multicolumn{1}{c|}{1000} &0.016 & 32.47 & \multicolumn{1}{c|}{95.00} & 0.015 & 42.44 & \multicolumn{1}{c|}{95.20} &0.014 &46.58 & \multicolumn{1}{c}{94.40}\\
			 \multicolumn{1}{c|}{1500} &0.011 & 36.79 & \multicolumn{1}{c|}{ 96.00}& 0.010& 46.00 & \multicolumn{1}{c|}{95.60 } &0.010 & 44.31 & \multicolumn{1}{c}{95.60 }\\ \hline
			 
			\multicolumn{1}{c|}{ }& \multicolumn{3}{c|}{ } & \multicolumn{3}{c|}{{\small \textbf{SBM}}}&\multicolumn{3}{c}{ } \\
			\multicolumn{1}{c|}{500} &0.025 & 29.48 & \multicolumn{1}{c|}{95.60}&0.021& 40.29 & \multicolumn{1}{c|}{95.80} &0.020 & 52.91 & \multicolumn{1}{c}{95.00}\\
			\multicolumn{1}{c|}{1000}  & 0.016 & 38.31 & \multicolumn{1}{c|}{96.00}& 0.014 & 42.85 & \multicolumn{1}{c|}{95.60 } &0.014 & 44.51 & \multicolumn{1}{c}{ 95.00}\\
			\multicolumn{1}{c|}{1500}  &0.012 & 37.74 & \multicolumn{1}{c|}{ 95.00}& 0.012& 42.07 & \multicolumn{1}{c|}{ 95.00} &0.012 & 43.40 & \multicolumn{1}{c}{94.00 }\\ \hline
			
			\multicolumn{1}{c|}{ }& \multicolumn{3}{c|}{ } & \multicolumn{3}{c|}{{\small \textbf{LSM}}}&\multicolumn{3}{c}{ } \\
			\multicolumn{1}{c|}{500}  &0.035& 26.06 & \multicolumn{1}{c|}{93.40}& 0.034 &24.16 & \multicolumn{1}{c|}{95.20} &0.031 & 24.14 & \multicolumn{1}{c}{93.75}\\
			\multicolumn{1}{c|}{1000}  &0.026 & 16.96 & \multicolumn{1}{c|}{ 94.20}& 0.023 & 17.42 & \multicolumn{1}{c|}{94.00 } & 0.022 & 17.86 & \multicolumn{1}{c}{95.03 }\\
			\multicolumn{1}{c|}{1500}  &0.020& 12.42 & \multicolumn{1}{c|}{ 93.72}&0.020&15.17& \multicolumn{1}{c|}{ 94.13 } &0.021 & 14.73 & \multicolumn{1}{c}{ 93.00}\\ \hline
	        \bottomrule
		\end{tabular}}
	\label{table:large p}
\end{table}

\begin{figure}[tb]
	\centering
	\includegraphics[width=6in]{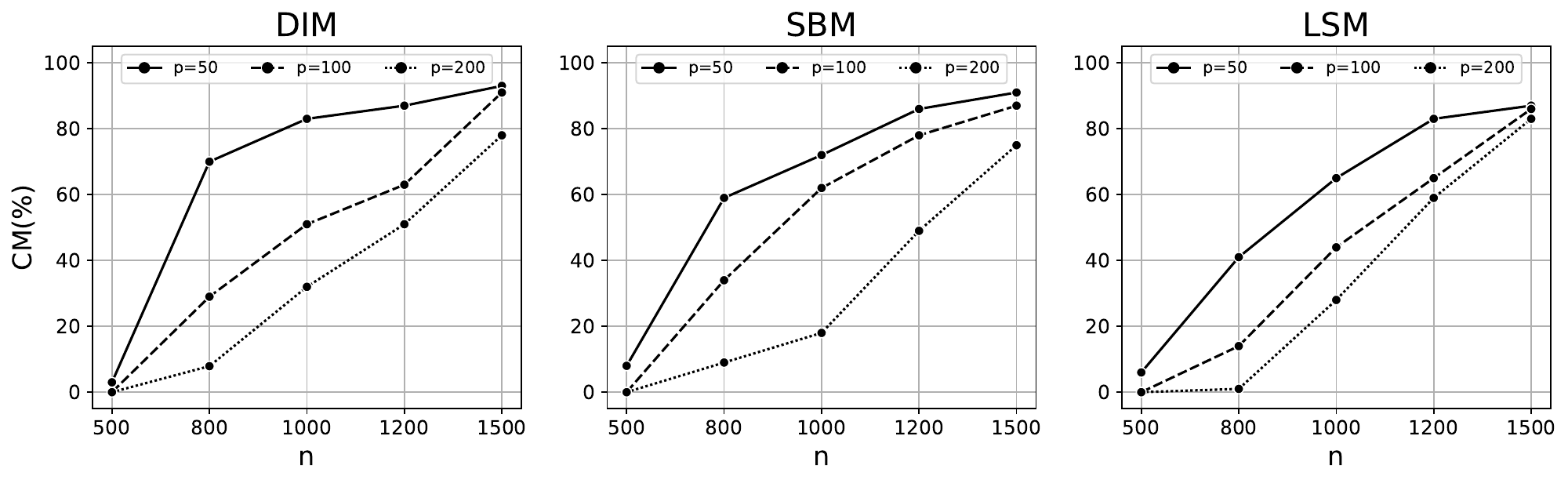}
	\caption{The CM values with $q=20$ and $d=3$. Different panels correspond to different network structures:  DIM (the left), SBM (the middle), and LSM (the right). For a given panel, different lines correspond to different feature dimensions with $p=50$ (solid), $100$ (dashed) and $200$ (dotted), respectively.} 
	\label{fig:CM}
\end{figure}

Lastly, we study the model selection results. Following Section \ref{subsec_Shrink}, we compute the SCAD estimators and use the BIC method to select the optimal tuning parameter $\lambda$. In this case, we randomly replicate the experiment for a total of $R=100$ times for each $(n,p,q,d,W)$ specification. Let $\wh{\mS}_{(j),\wh{\lambda}_{(j),\text{BIC}}}^{(r)}$ represent one particular model set obtained in the $r$-th replication ($1\leq r\leq R$). Define the percentage of experiments with correctly identified true models (CM) as 
\begin{equation*}
	\text{CM} =\dfrac{1}{R}\sum_{r=1}^RI\Big(\wh{\mS}_{(j),\wh{\lambda}_{(j),\text{BIC}}}^{(r)}=\mS_{(j),T}, \text{ for every }1\leq j\leq p\Big) \times 100\%.
\end{equation*}
This provides a uniform criterion for assessing model selection accuracy. Since the simulation results are qualitatively similar, we only report here case with $(q,d)=(20,3)$ in Figure \ref{fig:CM}. By Figure \ref{fig:CM}, we find that CM values converge to 100\% rapidly as the sample size $n$ increases. This suggests that $\wh{\mS}_{(j),\wh{\lambda}_{(j),\text{BIC}}}^{(r)}$ is uniformly consistent for recovering $\mS_{(j),T}$, which is in line with our theoretical findings in Theorem \ref{thm5}.

\csubsection{A Real Data Example}\label{subsec_real}

To demonstrate the practical applications of the proposed FSAR model, we present here a case study. Specifically, we consider an urban statistics dataset collected from Urban Statistical Yearbook 2019 of China, which is published by National Bureau of Statistics (\url{http://www.stats.gov.cn/sj/ndsj/}). The full dataset contains a total of 287 nodes, with each node representing a city. For each node (i.e., city), we collect a total of 112 macroeconomic indicators from 2019. Those indicators provide detailed information on city-level statistics. However, some indicators suffer from a large proportion (more than 15\%) of missing values and are then omitted for the subsequent analysis. For the remaining 50 indicators, the proportion of missing values does not exceed 5\%. Details of these 50 indicators are provided in Table 2 of Appendix D. For these 50 indicators, the neighbor year interpolation method of \cite{lunardi2018interpolation} is employed to impute the missing values. Subsequently, these completed indicators are then log-transformed and standardized to have mean 0 and variance 1. This leads to a final high-dimensional dataset $Y=(Y_{ij})\in\mR^{n\times p}$ with $n=287$ and $p=50$. A spatial weight matrix $W=(w_{i_1i_2})\in\mR^{n\times n}$ is then constructed based on geographical locations \citep{lee2010estimation}. We next fit an FSAR model to this dataset. Note that each node (i.e., city) is spatially connected with others through $W$. The associated spatial correlations $\rho_j$s reflect the spatial spillover effects among cities \citep{zhou2023spatial}.
\begin{figure}[tb]
	\centering
	\includegraphics[width=4in]{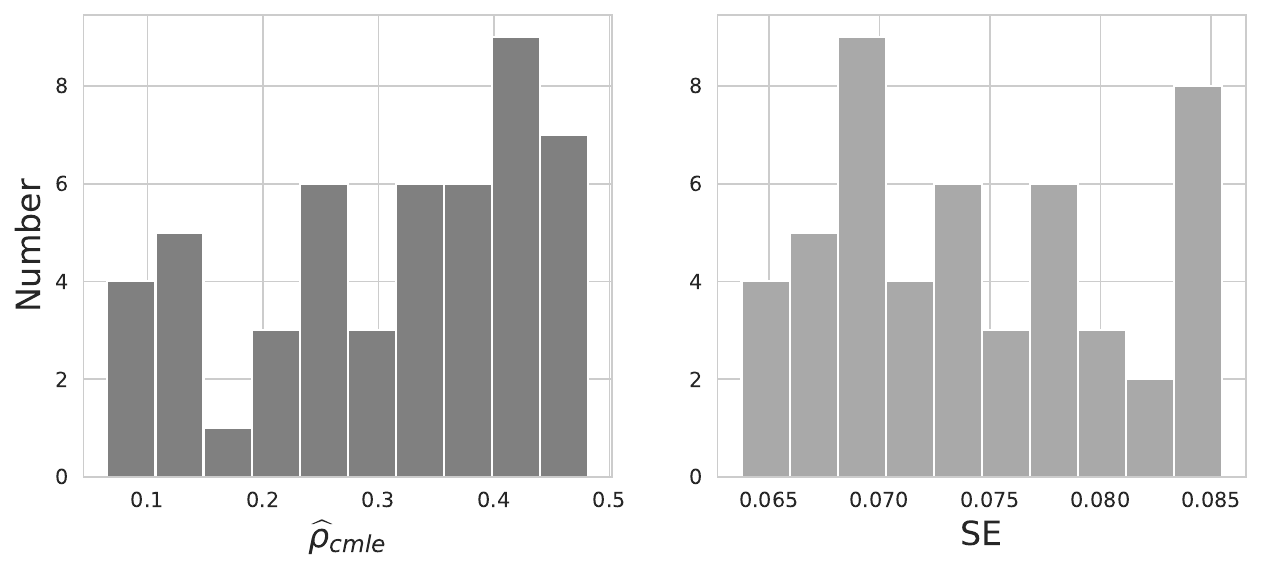}
	\caption{The histogram of the CMLE $\wh{\rho}_{j,\tcmle}$s (the left panel) and the histogram of the estimated SE (the right panel).}
	\label{fig:rhocmle}
\end{figure}

\begin{figure}[tb]
	\centering
	\includegraphics[width=3in]{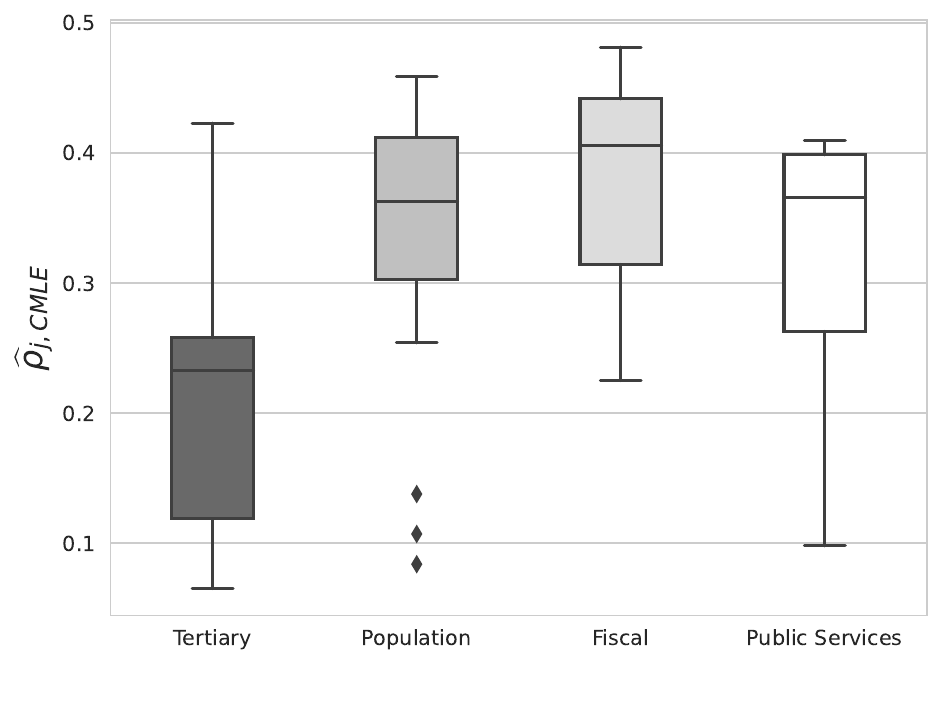}
	\caption{The boxplots of the CMLE $\wh{\rho}_{j,\tcmle}$s for four categories. From left to right: Tertiary (tertiary industry), Population (labor and population development), Fiscal (fiscal and financial resources), and Public Services (infrastructure and public services).}
	\label{fig:rhocmle_4c}
\end{figure}

We start with computing the CMLE  as an important initial estimator. The histograms of those estimators and their standard errors (SE) are then plotted in the left and right panel of Figure \ref{fig:rhocmle}, respectively. We find that the resulting estimates $\wh{\rho}_{j,\tcmle}s$ varies greatly, ranging from 0.05 to 0.50 with estimated SEs ranging from 0.06 to 0.09. It is then of interest to understand the reason behind such considerable variation. To this end, we classify the 50 indicators into a total of four groups. They are, respectively, (1) tertiary industry related indicators, capturing the development of services and knowledge-based sectors such as accommodation and food services,and wholesale and retail trade \citep{kenessey1987primary}; (2) labor and population development related indicators, reflecting the dynamics of urban labor supply, employment structure, and demographic shifts \citep{fujita2001spatial}; (3) fiscal and financial resources related indicators, indicating the capacity of local governments to mobilize and allocate financial resources, the strength of local fiscal institutions, and the accessibility of financial services \citep{gyourko1991structure}; and (4) infrastructure and public services related indicators, measuring the regional capacity to provide essential physical and social infrastructure \citep{demurger2001infrastructure}. The group sizes are 14, 16, 13, and 7, respectively. The CMLEs of each group are then boxplotted in Figure \ref{fig:rhocmle_4c}. We find that the spatial spillover effects of fiscal and financial resources are the strongest, with the largest $\wh{\rho}_{j,\tcmle}$s on average. This result aligns well with empirical findings in the urban and regional economics literature \citep{gyourko1991structure,auerbach2020local}.  
It is also noteworthy that the spatial spillover effects of tertiary industry are the weakest on average. 
Practically, this pattern may be explained by the relatively localized nature of the service-oriented economic activities \citep{kenessey1987primary,yin2022spatial}. 

\begin{figure}[tb]
	\centering
	\includegraphics[width=4in]{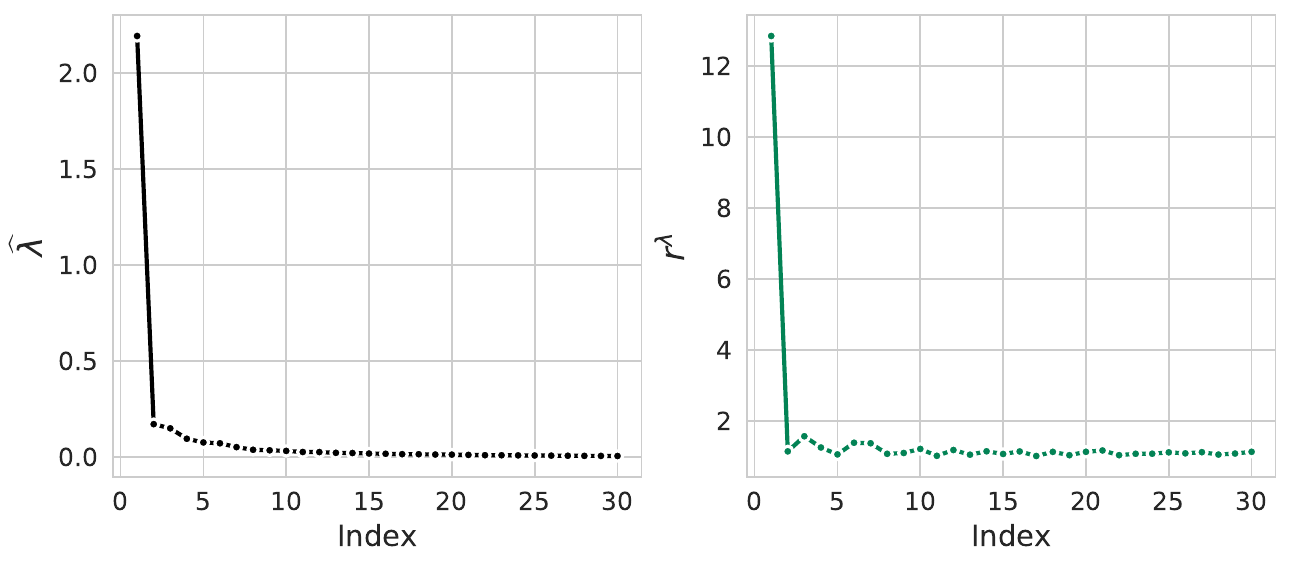}
	\caption{The top 30 estimated eigenvalues of $\wh{\mathbb{E}}$ (the left panel) and the top 30 eigenvalue ratios (the right panel).}
	\label{fig:eigen}
\end{figure}

Next, we need to decide the factor dimension. To this end, we compute $\wh{\varepsilon}_{i} = (\wh{\varepsilon}_{ij})\in\mR^p$ for each city $i$. Then, we compute the eigenvalues ($\wh{\lambda}_1\geq\cdots\geq \wh{\lambda}_p$) of the covariance matrix of  $\wh{\mathbb{E}}=\big(\wh{\varepsilon}_{ij}\big)\in\mR^{n\times p}$. The top 30 eigenvalues are then plotted in the left panel of Figure \ref{fig:eigen}. It seems that the first eigenvalue is notably larger than others. Following \cite{luo2008contour} and \cite{lam2012factor}, we calculate the eigenvalue ratio statistic as $r_j^{\lambda}=\wh{\lambda}_j/\wh{\lambda}_{j+1}$ with $1\leq j\leq p-1$. These values are then plotted in the right panel of Figure \ref{fig:eigen}, which provides strong evidence for the existence of a one-dimensional factor structure. This finding is not totally surprising, as these 50 indicators are all macroeconomics related. Therefore, they are heavily correlated with the overall macroeconomics status (i.e., one common factor) of the target city. This makes the underlying factor structure of $\wh{\varepsilon}_i$ relatively simple. Similar low-dimensional factor structures are also often observed in empirical macroeconomics literature \citep{bai2002determining,bernanke2005measuring}.

\begin{figure}[tb]
	\centering
	\includegraphics[width=4in]{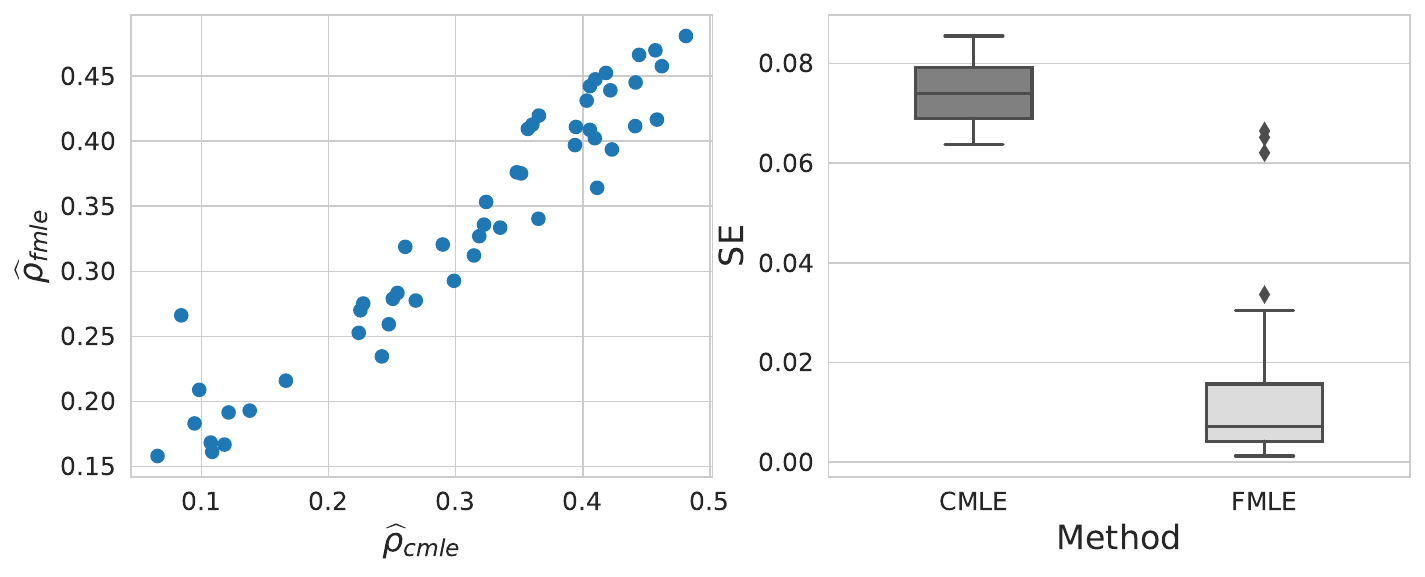}
	\caption{The scatter plot of the FMLE $\wh{\rho}_{\tfmle}$ and the associated CMLE $\wh{\rho}_{\tcmle}$ (the left panel). The boxplots of their standard errors (the right panel).}
	\label{fig:fmle}
\end{figure}

Lastly, we apply the proposed factor estimation method in Section~\ref{subsec_FactorDP} and obtain the estimated latent factor $\wh{Z}_i$ for each city $i$. The choice of the projection matrix is similar to that in simulation studies. Next, we compute the FMLE for every $\rho_j$ ($1\leq j\leq p$). The resulting FMLE estimates, along with the initial CMLE estimates, are then plotted in the left panel of Figure \ref{fig:fmle}.  We find that $\wh{\rho}_{\tfmle}$ is in line with $\wh{\rho}_{\tcmle}$. Moreover, their standard errors are boxplotted in the right panel of Figure \ref{fig:fmle}. We find that the SEs of FMLE are considerably smaller than those of CMLE. Our estimation results reveal that there exists the significant spatial correlation in various macroeconomic indicators among these cities. Specifically, the largest spatial spillover effect is detected for revenue in the gross regional product (GRP) growth rate with $\wh{\rho}_{1,\tfmle} = 0.48$. In contrast, the smallest spatial spillover effect is detected for persons employed in culture, sports and entertainment with $\wh{\rho}_{37,\tfmle} = 0.16$.  

\csection{CONCLUDING REMARKS}\label{sec_conclu}

In this work, we study the problem of spatial autoregressive modeling for network data with high-dimensional responses and covariates. The key contribution lies in the development of a flexible factor-augmented spatial autoregressive (FSAR) model that accommodates both high-dimensionality and complex cross-sectional dependence across response variables. To conclude this article, we discuss here several interesting topics for future research. First, it is worth noting that the FSAR model requires the high-dimensional responses to be continuous. Then how to relax this continuity assumption is an interesting direction for future exploration. Second, it is assumed that the dimension of the latent factors is fixed. How to allow for a diverging number of latent factors should be another intriguing topic for the future study \citep{fan2008high}. Third, for the real urban statistics dataset analysis, the cross-response spillover effects (e.g., the cross effect of retail sales and household wealth) are not explicitly characterized. Developing novel tools for better interpretation is also worth pursuing.

\textbf{Supplementary Materials.} Appendices A--C provide the proofs of all theoretical results and some useful lemmas, and Appendix D contains the supplementary table. The code is publicly available on GitHub at \url{https://github.com/Shi12056/FactorSAR.git}.

\textbf{Acknowledgments.} The authors are very grateful to the editor, the associate editor, and referees for their constructive comments and suggestions, which greatly improved the quality of this paper.

\textbf{Disclosure Statement.} The authors report there are no competing interests to declare.

\textbf{Funding.} Xuening Zhu's research is supported by the National Natural Science Foundation of China (nos. 72222009, 71991472, 12331009), MOE Laboratory for National Development and Intelligent Governance, Fudan University. The research of Jing Zhou is partially supported by the National Natural Science Foundation of China (Nos. 72171226, 11971504) and the National Statistical Science Research Project (No.2023LD008). Hansheng Wang’s research is partially supported by the National Natural Science Foundation of China (Nos. 12271012).

\bibliographystyle{asa}
\bibliography{reference}

%

\end{CJK}
\end{document}